\documentclass[12pt,aps,roupedaddress,aa]{revtex4}

\usepackage{color}
\usepackage{latexsym}

\usepackage{amsmath}
\usepackage{graphicx}

\usepackage{dcolumn}

\input epsf

\def\lesssim{\mathrel{\hbox{\rlap{\hbox
{\lower4pt\hbox{$\sim$}}}\hbox{$<$}}}}
 
\def\gtrsim{\mathrel{\hbox{\rlap{\hbox
{\lower4pt\hbox{$\sim$}}}\hbox{$>$}}}}

\begin{document}

\title{Numerical simulation of small perturbation on an accretion disk due 
to the collision of a star with the disk near the black hole }

% repeat the \author\address pair as needed
\author{Orhan D\"{o}nmez}

\address{Nigde University Faculty of Art and Science, 
Physics Department, Nigde, Turkey} 

%\address{Department of Physics, Drexel University, Philadelphia 19104}
%\address{NASA Goddard Space Flight Center, Greenbelt, MD 20771}

\date{\today}

\begin{abstract}

In this paper, perturbations of an accretion disk by a star orbiting around a black hole
are studied. We report on a numerical experiment, which has been carried out by using 
a parallel-machine code originally developed by D\"{o}nmez (2004). 
An initially steady state accretion disk near a non-rotating 
(Schwarzschild) black hole interacts with a "star", modeled as an initially circular 
region of increased density. Part of the disk is affected by the interaction. 
In some cases, a gap develops and shock wave propagates through the disk. 
We follow the evolution for order 
of one dynamical period and we show how the non-axisymetric density perturbation further 
evolves and moves downwards where the material of the disk and the star become 
eventually accreted onto the central body.

When the star perturbs the steady state accretion disk, 
the disk around the black hole is destroyed by the effect of perturbation. 
The perturbed accretion disk creates a shock wave during the evolution 
and it  loses angular momentum when the gas hits on the shock waves.
Colliding gas with the shock wave is the one of the basic  mechanism of emitting 
the $X-$rays in the accretion disk. The series of supernovae occurring 
in the inner disk could entirely destroy the disk in that region which leaves
a more massive black hole behind, at the center of galaxies. 

\end{abstract}

\maketitle

\section{INTRODUCTION}

It is widely accepted that the supermassive black holes (SBH) are located in the 
center of active galaxies and quasars (Begelman et al. 1984). The mass 
of SBH is 
usually estimated to be in the range $M \approx  10^6 - 10^{8} M_{\odot}$.
Anomalous energy output of active galactic nuclei (AGN) may result
from an accretion process. The accretion process involves matter which
is attracted from the surroundings of AGN or comes from tidily disrupted 
stars passing too near by the SBH. The accretion disk is formed and 
eventually the matter falls into the black hole. The best evidence comes 
from studies of the central surface brightness of the nuclei, stellar 
velocity dispersion, spatial distribution of stars and chaotic time 
variations in $X-$ray emission (Young et al. 1978, Dressler et al. 1988, 
Dressler et al. 1990, Artymowicz et. al. 1993, Kato 2001). General 
relativistic effect is  important consequences for the axial symmetry, 
stability and some of the phenomena of accretion disks. Artymowicz et. al. 1993
estimates efficiency for rapid metallicity enrichment of the AGN and quasars based on 
star-disk interaction. The accretion disk around the massive black hole involves capture of low-mass
stars from the host galaxy's nucleus by the accretion disk orbiting around the massive black hole.

The dissipate interaction of star with accretion disk and main star was 
considered by different authors. The tidal interaction of star with 
central object, the massive black hole, was discussed by 
Hills (1975) and Artymowicz et. al. (1993). The interaction of star with 
an accretion disk was considered in the works by Vilkoviskij et al. (1981),Vilkoviskij (1983),
Zentsova (1983), and Subr et. al. (2004). The gradual orbital decay of stellar  
trajectories is studied by Subr et. al. (2004). They have shown that how the stars become 
gradually flattened toward the accretion disk and identified  parameters affecting the forming
of structure during this process. The dynamical evolution of relativistic
star cluster around  a massive Kerr black hole with an accretion disk was examined by Rauch (1995).
They have looked at the regime where the black hole dominates the potential
and star-disk interaction. They have shown that there is an astrophysical
plausible regime in which an accretion disk around the black hole in the AGN 
would dominate the dynamical evolution of star cluster through the star-disk
interaction. Rauch (1995) and Artymowicz et. al. (1993) also declared that a series of 
supernovae occurring 
in the inner disk could entirely destroy the disk in that region which leaves
a more massive black hole or neutron star behind. Ivanov et al. (1998) 
considered the hydrodynamical effects
resulting from a collision among a supermassive black hole and an accretion
disk orbiting another massive black hole which is bigger than the other. They
studied these effects both analytically and numerically. They declared that 
the black hole-disk collision results in significant changes on disk structure, and 
gravity of the moving black hole induces a converging gas flow in the disk. 
This converging gas helps to form a shock waves on the accretion disk.
Subr et al. (1999) consider eccentric orbit intersecting disk once or twice 
per each revolution and they solve the equations for oscillating elements. 
They found the thermal radii of the orbits and time needed to bring the 
orbit into the disk plane as function of initial parameters, and they 
showed  that previous simplified estimates, which is derived for the 
case of low eccentricity, remain valid within factor of two. The star disk 
interaction and gravitating disk around the black hole are reviewed within 
different approaches and with emphasis on the role of disk gravity by 
Karas et al. (2004).

Theoretical prediction of star-disk interaction near a massive black
hole is examined by Syer et al. (1991). Two different scenarios could happen
when stars are orbiting on their eccentric orbit. The stars can pass
through  the accretion disk and create a hole or can be ground down
by gravitational forces. The stars in  galaxies could be 
ground  down into a short period orbit during the star-disk
interaction around  the black hole. Any stars on a highly eccentric
orbit, which are close to a black hole,  would lose their potential
energy. It may take a long time to be capture stars and grind down 
into an accretion disk. 
In Syer et al. (1991), they have considered mechanisms whereby stars 
can get on to orbits within $q_{coll} = GM/v^{2}_{*}$, where $v_*$ is 
the escape velocity from the star, $M$ is the mass of the black hole
 and $q_{coll}$ is major axis, without being 
destroyed. It might be thought that this could happen by a tidal capture
process. if the stars get closer to the black hole  than $q_t = 
(M/m_*)^{1/3} r_*$, they are tidily disrupted. A star captured into 
an orbit with pericenter at $3q_t$ would undergo further tidal dissipations.
The greatly increased geometric area  and reduced binding energy suggests that
such stars should be susceptible to large-scale stripping of mass on collision 
with the disk (Zurek et al. 1991). Within some critical radius $R_{crit}$, 
collisions will then destroy the star and unbind the envelop mass within the 
red giant lifetime. The value of $R_{crit}$ will depend on the extent and
properties of the disk as well as velocity and structure of star. If 
$R_{crit}$ and the central stellar density are sufficiently large, this
mechanism could then provide an important source of fuel for the central 
black hole (Armitage et al. 1996).

Different sorts of capture have also been suggested as formation mechanisms 
for binary stars. In these scenarios,  some sort of
process, which can dissipate kinetic energy, exist so that two unbound stars 
can become bound during a close encounter. These can
be  tidal capture where two stars through a close passage 
excite tidal modes in one another which absorb much
of their relative kinetic energy and dynamic capture where three stars 
pass close together, exchanging energy so that one is
ejected with a large part of the kinetic energy leaving the two other 
stars bound. The problems about these two models are that
they require a high density of stars to make such close encounters 
statistically significant and thus probably can't form the
majority of the overall binary population. It should be noted however 
that tidal capture is still the leading mechanism for
formation of X-ray binaries in dense clusters 

In this paper, we numerically model the interaction of the accretion disk with star 
using general relativistic hydrodynamical code and assumed that 
the SBH forms such a binary system with a low mass star in an orbit 
inclined  with respect to the plane of an accretion disk.  The disk is 
thin and at the equatorial plane of the black hole but the model can be 
generalized to a more complicated model. We ignored the self gravity 
among the matter of accretion disk. The gravity between the black hole
and accretion disk is set up by the Schwarzschild black hole.

\section{FORMULATION}
\label{formulation}

The General Relativistic Hydrodynamic (GRH) equations in  Refs.
Donat et al. (1998), Font et al. (2000), and D\"{o}nmez (2004), written in the standard
covariant form, consist  of the local conservation laws of the
stress-energy tensor $T^{\mu \nu }$  and the matter current density $
J^\mu$:

\begin{eqnarray}
\bigtriangledown_\mu T^{\mu \nu} = 0 ,\;\;\;\;\; 
\bigtriangledown_\mu J^\mu = 0.
\label{covariant derivative}
\end{eqnarray}

\noindent
Greek indices run from $0$ to $3$, Latin indices from $1$ to $3$, and units 
in which the speed of light $c = 1$ are used.

Defining the characteristic waves of the general
relativistic hydrodynamical equations is not trivial with imperfect
fluid stress-energy tensor. We neglect the viscosity and heat
conduction effects.  This defines the  perfect fluid
stress-energy tensor. We use this stress-energy tensor to derive the
hydrodynamical equations. With this 
perfect fluid stress-energy tensor, we can solve some problems which
are  solved by the Newtonian hydrodynamics with viscosity, such as
those involving angular momentum transport and shock waves on an
accretion disk, etc. Entropy for 
perfect fluid is conserved along the fluid lines. The stress energy tensor
for a perfect fluid is given as

\begin{equation}
T^{\mu \nu} = \rho h u^\mu u^\nu + P g^{\mu \nu}.
\label{des 7}
\end{equation}

\noindent
Here  $h$ is the specific enthalpy and 4-velocity $u^{\mu}$ which may vary from 
event to event. It exhibits a density of mass $\rho$ and isotropic pressure $P$ 
in the rest frame of each fluid element
The equation of state might have  the 
functional form $P = P(\rho, \epsilon)$. The perfect
gas equation of state, 

\begin{equation}
P = (\Gamma -1 ) \rho \epsilon.
\label{flux split21}
\end{equation}

The conservation laws in the form given in Eq.(\ref{covariant
derivative}) are not suitable for 
the use in advanced numerical schemes. To carry out numerical
hydrodynamic evolutions such as those reported in Font et al. (2000), and 
D\"{o}nmez (2004) and to
use high resolution shock capturing schemes, the hydrodynamic equations 
after the 3+1 split
must be written as a hyperbolic system of first order flux
conservative equations. We write Eq.(\ref{covariant derivative}) in
terms of coordinate derivatives, using the coordinates ($x^0 = t, x^1,
x^2, x^3$). Eq.(\ref{covariant derivative}) is projected onto the
basis $\lbrace n^\mu, (\frac{\partial}{\partial x^i})^\mu \rbrace$,
where $n^\mu$ is a unit timelike vector normal to a given
hypersurface. After a straightforward calculation,
we get (D\"{o}nmez (2004)),

\begin{equation}
\partial_t \vec{U} + \partial_i \vec{F}^i = \vec{S}. 
\label{desired equation}
\end{equation} 

\noindent
This basic step serves to identify the
set of unknowns, the vector of conserved quantities $\vec{U}$, and
their corresponding fluxes $\vec{F}(\vec{U})$. With the
equations in conservation form, almost every high
resolution method devised to solve hyperbolic systems of conservation
laws can be extended to GRH.

The evolved state vector $\vec{U}$ consists of  the conservative
variables $(D, S_j, \tau)$ which are conserved variables for density,
momentum and energy respectively; in terms of the
primitive variables $(\rho, v^i, \epsilon)$, this becomes

\begin{equation}
\vec{U} = \left( 
\begin{array}{c} 
D \\
S_j \\
\tau 
\end{array} 
\right) = \left( 
\begin{array}{c}
\sqrt{\gamma} W \rho \\
\sqrt{\gamma}\rho h W^2 v_j \\
\sqrt{\gamma} (\rho h W^2 - P - W \rho) 
\end{array} 
\right). 
\label{matrix form of conserved quantities}
\end{equation}

\noindent
Here $\gamma$ is the determinant of the 3-metric $\gamma_{ij}$, $v_j$
is the fluid 3-velocity, and W is the Lorentz factor.

\noindent 
The flux vectors $\vec{F^i}$ are,

\begin{equation}
\vec{F}^i =  \left( \begin{array}{c}
\alpha (v^i - \frac{1}{\alpha} \beta^i) D \\
\alpha \lbrace (v^i - \frac{1}{\alpha} \beta^i) S_j + \sqrt{\gamma} P
\delta ^{i}_{j} \rbrace    \\
\alpha \lbrace (v^i - \frac{1}{\alpha} \beta^i) \tau + \sqrt{\gamma}
v^i P \rbrace \end{array} \right). 
\label{matrix form of Flux vector}
\end{equation}

\noindent
The spatial components of the 4-velocity $u^i$ are related to the
3-velocity by the following formula: $u^i = W (v^i - \beta^i /
\alpha)$. $\alpha$ 
and $\beta^i$ are the lapse function and the shift vector of the
spacetime respectively. The source vector $\vec{S}$ for Schwarzschild coordinate is given by
D\"{o}nmez (2004)

\noindent

\begin{eqnarray}
\vec{S} =  \left( 
\begin{array}{c}
0 \\
\frac{1}{2} \alpha \sqrt{\gamma} (T^{tt} \partial_{r} g_{tt} + T^{rr} \partial_{r} g_{rr} + \nonumber 
T^{\theta \theta} \partial_{r} g_{\theta \theta}  + T^{\phi \phi}
\partial_r g_{\phi \phi}) \\
\frac{1}{2} \alpha \sqrt{\gamma} T^{\phi \phi}
\partial_{\theta} g_{\phi \phi} \\
0 \\
\alpha \sqrt{\gamma} (T^{r t}
\partial_{r} \alpha - \alpha T^{r t} g^{tt} \partial_{r} g_{tt})
\end{array} 
\right). 
\label{matrix form of source vector3}
\end{eqnarray}

More detailed information related with numerical solution of GRH equations using high 
resolution shock capturing schemes in $3D$ are given in our paper D\"{o}nmez (2004). 
This reference explains detail description of numerical solutions of SRH equations, 
Adaptive-Mesh Refinement, high resolution shock capturing method used, 
and solution of GRH equation using  Schwarzschild coordinate as a source term. The code is 
constructed for general spacetime metric with lapse function and shift vector and is fully 
parallelized to make optimum use of supercomputers which is necessary to achieve the 
numerical modeling of real astrophysical problems, such as coalescing of the compact binaries
and accretion disk around the compact objects.

%%%%%%%%%%%%%%%%%%%%%%%%%%%%%%%%%%%%%%%%%%%%%%%%%%%%%%%%%%%%%%%%%%%
%%%%%%%%%%%%%%%%%%%%%%%%%%%%%%%%%%%%%%%%%%%%%%%%%%%%%%%%%%%%%%%%%%%%
%%%%%%%%%%%%%%%%%%%%%%%%%%%%%%%%%%%%%%%%%%%%%%%%%%%%%%%%%%%%%%%%%%%%

\section{The Physical Model  and Constructing the Initial Accretion Disk}
\label{The Physical Model  and Constructing the Inital Accretion Disk}

The structure and the dynamics of the accretion disk remain quite mysterious. 
The disk is not directly observable because the resolution of current telescopes is 
still insufficient. It is primarily studied at short wavelengths (UV, X and gamma rays).
But short wavelength spectra give information only on the internal regions of the disk 
(scale of the micro-parsec), very close to the black hole. One suspects that at these 
distances, the mass of the disk (generally regarded as small) starts to play a role 
on its own dynamics, and thus on its evolution and its structure. One then expects very 
particular effects, such as the generation of gravitational instabilities (spiral waves)

The spiral structure on an accretion disk is inevitable in case of unstable situations. 
To simulate real accretion disk structure, the larger disk is needed. Despite 
large body of numerical work, there has been relatively little progress in the 
quantitative measurement of  the effective transport of spiral shock in hydrodynamic 
simulations. This is due in large part to  the fact that it is easy to set up a 
hydrodynamic simulation of a Keplerian accretion disk, but it is very hard to do it 
accurately enough to measure to relatively small transport expected due to spiral wave. 
To define structure of disk, one needs high numerical accuracy, a large  range in radii, 
and  long simulations covering several orbital periods. In order to watch the  behavior 
of accretion disk and reduce the complexity, we restrict our attention to close to  the black 
hole and put apply some restrictions explained below .

\subsection{Method and Assumptions}
\label{Method and Assumptions}

To numerically model the accretion disk dynamics using the parallel-machine code, which
can handle strong shock cases, we have made some assumptions.
The computational model consists of two components: A central massive
black hole and associated accretion disk lying in the equatorial plane 
of the hole with captured star. We presumed that the disk around the black 
hole is in the steady state condition. To restrict our attention to the 
influence of star-disk interactions on the evolution, and to keep the 
problem numerically tractable, the following assumptions on the calculations
are made: 
i) the central black hole dominates the gravitational potential; 
ii) the accretion disk is thin and in steady state, which is created by 
numerical model; iii) the stars impact the disk supersonically;
iv) captured star-disk interactions dominate the evolution of 
the stellar orbit. The thin accretion disk models describes an advection-dominated 
flow in which the radiative cooling is inefficient and most of the dissipated 
energy is advected into the black hole.
In fact, the first constraint merely requires that the mass of the black 
hole be much larger then the mass of the cluster and disk. 
We neglect the self gravity. Outher boundary of numerical domain is chosen  maximum in radial
direction, $R_{max} =30M$, $M$ is mass of the black hole,  in where gravity of the  
central massive black hole is
too dominant  compared with self gravity of disk. The validity of 
the assumption that star-disk collisions is dominating  the evolution depend
on both the density of the disk  and the density of stars in the cluster.
The problem of disks around the black hole is discussed within the general relativistic
regime. Whereas general relativity is mainly relevant for the inner parts of disk
(within $100M$ Schwarzschild radius), Newtonian regime is adequate outside 
this radius. In the rest of the paper, distances in numerical calculations and time  
depend on the mass of the black hole $M$.

%%%%%%%%%%%%%%%%%%%%%%%%%%%%%%%%%%%%%%%%%%%%%%%%%%%%%%%%%%%%%%%%%%%%%%%%%%
\subsection{Modeling a steady state initial accretion disk}
\label{Modeling of Steady State Initial Disk}

The numerical approximation of captured star by disk in the
close binary system are modeled and evolved by solving the GRH 
equations and using fully GRH code (Ref. D\"{o}nmez 2004), run in two-dimensional 
spherical coordinates $(r,\phi)$ centered on the accretion disk, 
called equatorial plane. Theoretically, it is known that when the 
stars are orbiting on their eccentric orbit, they can be ground down or make
a hole in the accretion disk around the black hole. The modeling of the perturbation needs 
an accretion disk, which could be in steady 
state or not, around the black hole. After the star is captured by disk, the region where 
star falls down during the evolution may not be affected by the star. On 
the other hand, steady state accretion disk can be used in the hole case 
which is created when stars are rotating in their eccentric orbit. 
Since there is no self gravity in our numerical code, the mass of the accretion disk 
does not effect the solutions. 

In order to simulate these problems,
initial steady state accretion disk is  computed at the equatorial plane ($r-\phi$)
of the black hole in the domains $3M \leq r \leq 30M$ 
and $0 \leq \phi \leq 2\pi$. The $\phi$ boundaries are periodic and the inner and outer
radial boundaries are constructed to allow material to flow off the grid, called outflow, 
but not onto grid. This allows gas to fall into the black hole and out the computational domain 
when it reaches the inner
boundary and the outer boundary in radial direction, respectively. An accretion disk  is a 
structure formed by material falling into a gravitational source. 
Before applying any perturbation on the 
accretion disk, we start our analysis under the simplifying assumptions that the black hole has
has steady state accretion disk around and it is numerically crated 
around the black hole choosing appropriate initial values.

To create initial steady state accretion disk numerically, 
a sub-Keplerian velocity is chosen as an initial $\phi$ velocity of gas because the 
gravitational forces will be balanced by the pressure and centrifugal  
forces on the accretion disk. Our simulations were run on uniform grid using $256$ 
points in radial  and $128$ in angular distances. For stability reasons, time steps 
change with changing sound velocity and velocity of matter. While initial disk 
simulation ran for over 
$11000$ cycles, perturbation of accretion disk models ran for over $15 10^4$ cycles. 
To create steady state accretion disk from the numerical simulation, the initial 
values of variables are;

\begin{center}
\begin{figure}
\centerline{\epsfxsize=7.cm \epsfysize=7.cm
\epsffile{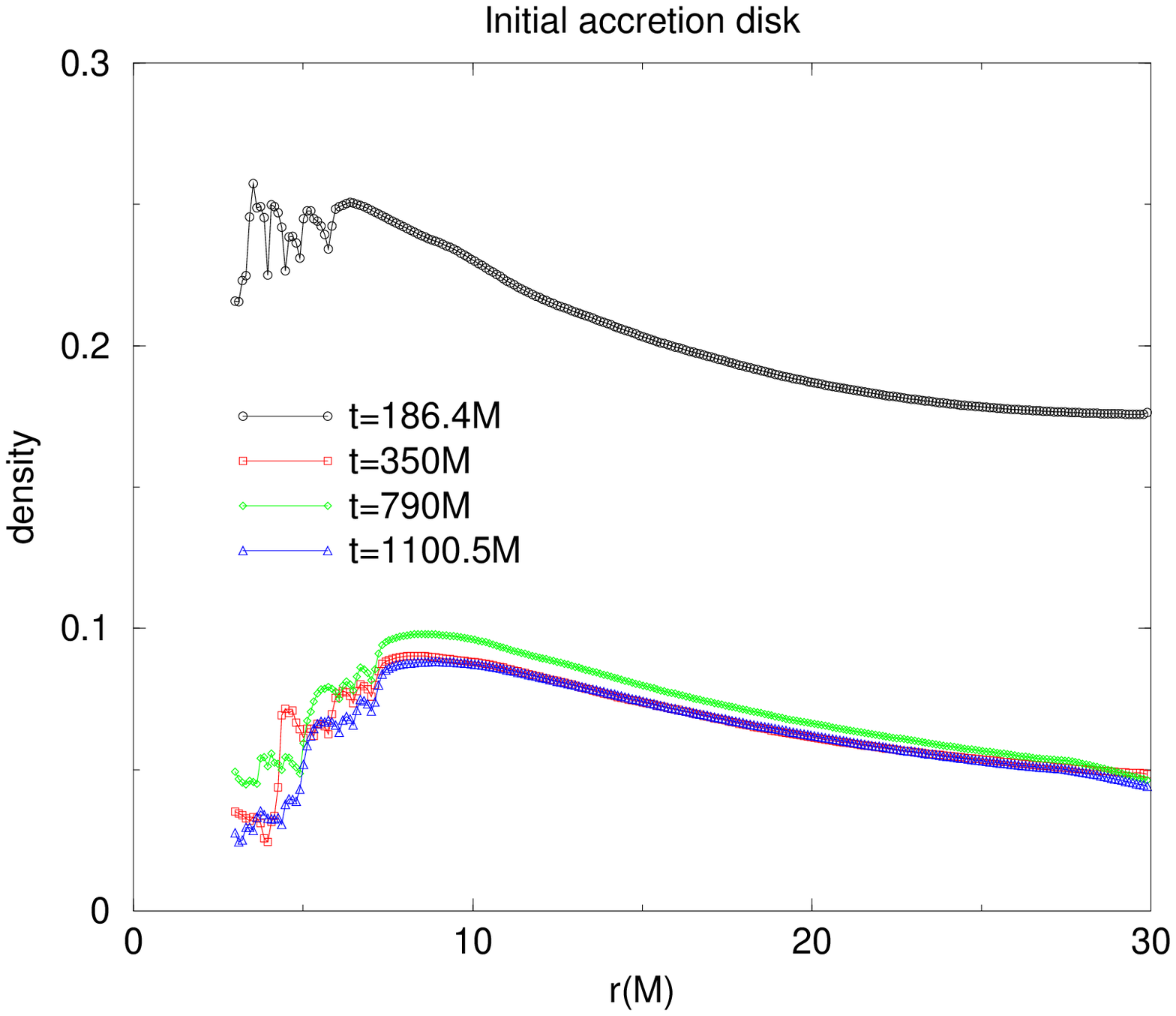}}
\centerline{\epsfxsize=7.cm \epsfysize=7.cm
\epsffile{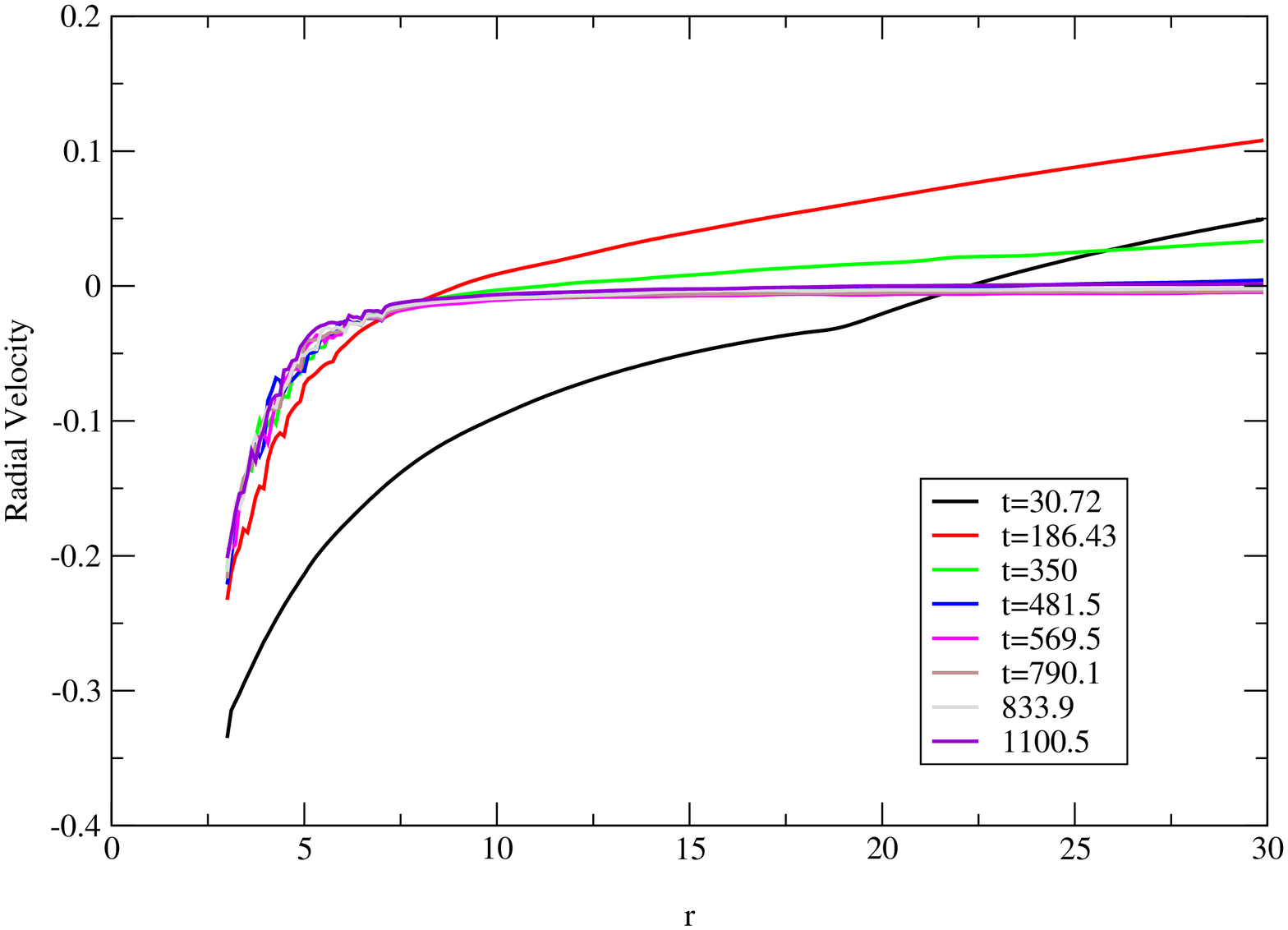}}
\caption{Creation of the initial accretion disk during the time evolution. Top panel: density of 
the initial accretion disk vs. radial coordinate. The accretion disk reaches to quasi-steady state at 
$t=1100.5M$. Bottom panel: radial velocity of gas during the creation of accretion disk. Radial 
velocity goes to zero outside the last stable orbit, around $6M$, when disk reaches to steady 
state. This disk is used as an initial condition during the rest of the paper in our numerical 
calculations.} 
\label{Inital_disk}
\end{figure}
\end{center}

\noindent
$\rho_{max} = 0.09$, $v^{\phi}_d = \sqrt{(M/r^3)}/\sqrt{(1 - (2 M)/r)} -
1/r^2$, $p=0.1$  and the radial velocity is zero outside the last stable
circular orbit. The numerical simulation is performed until a steady state accretion disk 
created, seen in Fig.\ref{Inital_disk}.  The material  in the quasi-steady state disk
would keep on going in circular orbits and there would be no release of energy  after 
the formation of the disk. Taking this as our initial condition, we then study the evolution
of star-disk and gap-disk interaction close to the black hole using fully GRH code.

%%%%%%%%%%%%%%%%%%%%%%%%%%%%%%%%%%%%%%%%%%%%%%%%%%%%%%%%%%%%%%%%%%%%%%%%%%
%%%%%%%%%%%%%%%%%%%%%%%%%%%%%%%%%%%%%%%%%%%%%%%%%%%%%%%%%%%%%%%%%%%%%%%%%%
%%%%%%%%%%%%%%%%%%%%%%%%%%%%%%%%%%%%%%%%%%%%%%%%%%%%%%%%%%%%%%%%%%%%%%%%%%

\section{Numerical Results}

The dynamical evolution of captured relativistic star clusters around
the Schwarzschild black hole with an accretion disk is examined in the 
regime where the black hole dominates the potential and star-disk
interactions dominate the evolution. To take into account the interactions among 
the satellite and the disk matter due to general relativistic effects which are constituted 
by central object, numerical solution of the fully general relativistic hydrodynamical 
equations is needed. Theoretical prediction of star-disk interaction near a massive black
hole is examined by Syer et al. (1991) and Subr et al. (1999). 
Two different scenarios can happen
when the star is orbiting throughout  its eccentric orbit. When the star can pass
through in the accretion disk, it creates a gap and a hole in some region or it can be ground down
by gravitational forces and in this manner, the accretion disk can be perturbed. 

Stars on highly eccentric orbits, which approach close to the black,  would lose 
their orbital (potential) energy and angular momentum due to dynamic 
friction. The rate of star-disk captures is controlled by
the rate of repopulating of orbits that intersect the disk. The captured stars 
at this scenario may give a perturbation and change dynamics of the accretion disk. 
Because there is considerable uncertainty in the actual value of the disk 
surface density and since the cluster consists of various stellar types ranging from compact 
stars to giants, there is also uncertainty in the star density that perturbs the accretion disk.
Here, we have chosen a certain size of star, defined in section \ref{Grounding Down Star}. 
and it would be possible to have this size of star, and it  can  be entirely embedded in the
disk when the inclination angle drops to critical value(Subr et. al. 2004). If the stellar mass
is large enough to create a gap close to the black hole, the star becomes tidally coupled with 
the disk and the matter of the disk  is sinked into the gap. Finally, 
these types of interaction  perturb the accretion disk and could  create non-linear phenomena.

This section describes two different GRH simulation varying the types
of perturbation, grounded down star  and gap on the steady state accretion disk using
actual size of the black hole and a small perturbation which is represented by add density 
on the initial disk.

%%%%%%%%%%%%%%%%%%%%%%%%%%%%%%%%%%%%%%%%%%%%%%%%%%%%%%%%%%%%%%%%%%%%%%%%%%
\subsection{Grounded Star}
\label{Grounding Down Star}

To model the grounded star onto an accretion disk numerically, a perturbation is 
applied on the steady state 
accretion disk created in \ref{Modeling of Steady State Initial Disk}. 
During the interaction of star with accretion disk and central object, the star 
can be grounded down by 
forces which are produced by gravitational, centrifugal, pressure and tidal forces 
between captured star and the black hole. 
The grounded star can be 
represented as a high density region on the accretion disk and initial 
condition for the star is given as follow: $\rho_s = 0.1$ located at  
$28M \leq r \leq 30M$ and $0 \leq \phi \leq 0.1$.
The density of star is \%5 higher than the steady state density of initial disk and
initial radial velocity, about expected velocity of star at the 
high density region, is given to star which
is $v^r = -0.01$ (minus means star is  falling to the black hole). 
The other variables are the same with initial steady state
accretion disk. This is the scenario which is called a disk perturbed by a captured star

\begin{center}
\begin{figure}
\centerline{\epsfxsize=12.cm \epsfysize=12.cm
\epsffile{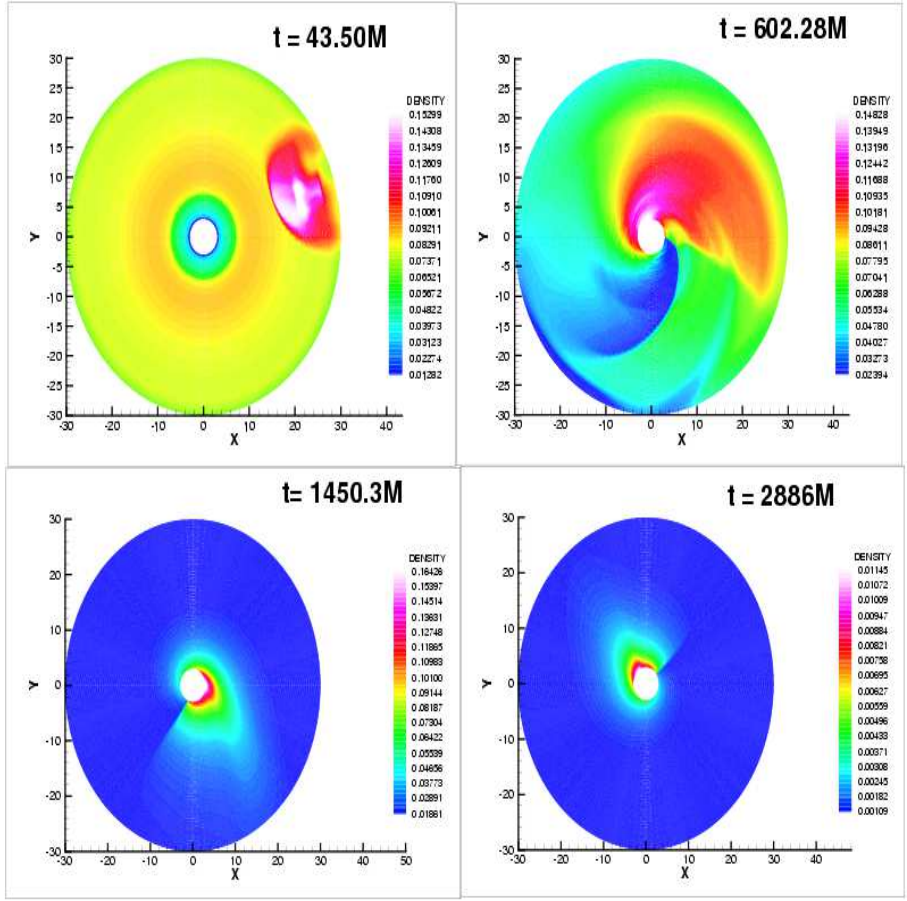}}
\caption{The plots represent four subsequent moments of evolution of grounded
star during the star disk interaction on the $r-\phi$ plane. The color corresponds the measure
of density $\rho$. The captured star is located close to the outer
boundary, seen at $t=43.5M$. The initial steady state disk is perturbed by star and disk 
matter  moves in an unstable manner, and heats up accretion disk. 
As a consequence of this phenomenon, a one-armed spiral shock wave is 
created at later times.
 } 
\label{falling star 1}
\end{figure}
\end{center}

\begin{center}
\begin{figure}
\centerline{\epsfxsize=12.cm \epsfysize=12.cm
\epsffile{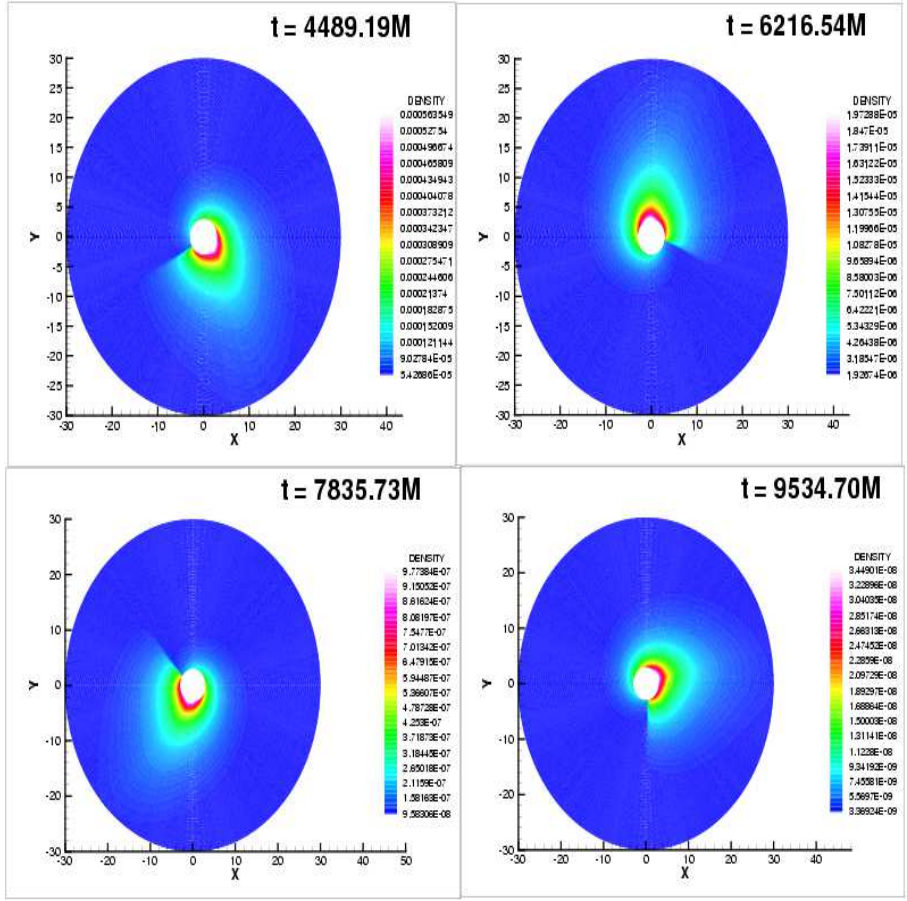}}
\caption{Same as Fig.\ref{falling star 1} but representation of density at later times. The
spiral shock rotates in a counterclockwise direction while accretion disk is rotating in clockwise
direction. } 
\label{falling star 2}
\end{figure}
\end{center}

Figs. \ref{falling star 1} and \ref{falling star 2} display the
density at different times on the equatorial plane.  The
grounded star which has a  higher density than accretion disk is moving toward
the black hole under the general relativistic, centrifugal, pressure and
hydrodynamical forces. The general relativistic future to the problem is given
by the Schwarzschild black hole. While the perturbed gas is falling into the black 
hole on the equatorial plane, it
creates one-armed spiral shock wave and shock wave propagates through the disk,
shown in Fig. \ref{falling star 1}. The spiral wave moves through the disk causing the star
and gas to clump up along the wave, called the density wave.
In the early time of simulation, the shock wave created on accretion disk
rotate in the same direction with the accretion disk and later, $t=1450.3M$, the
spiral shock is settled down and starts to counter-rotate. The dynamic of disk during
the later times  of simulation  is seen in
Fig. \ref{falling star 2}. At the spiral wave front, where the density is higher, 
star formation may be enhanced
in the vicinity of the black hole. Because the spiral wave travels through disk slower 
than the local Keplerian 
velocity, it carries negative angular momentum, and abrupt change or dissipation  at the shock 
decreases the angular momentum of the orbiting gas on disk. 

\begin{center}
\begin{figure}
\centerline{
\epsfxsize=12.cm \epsfysize=12.cm
\epsffile{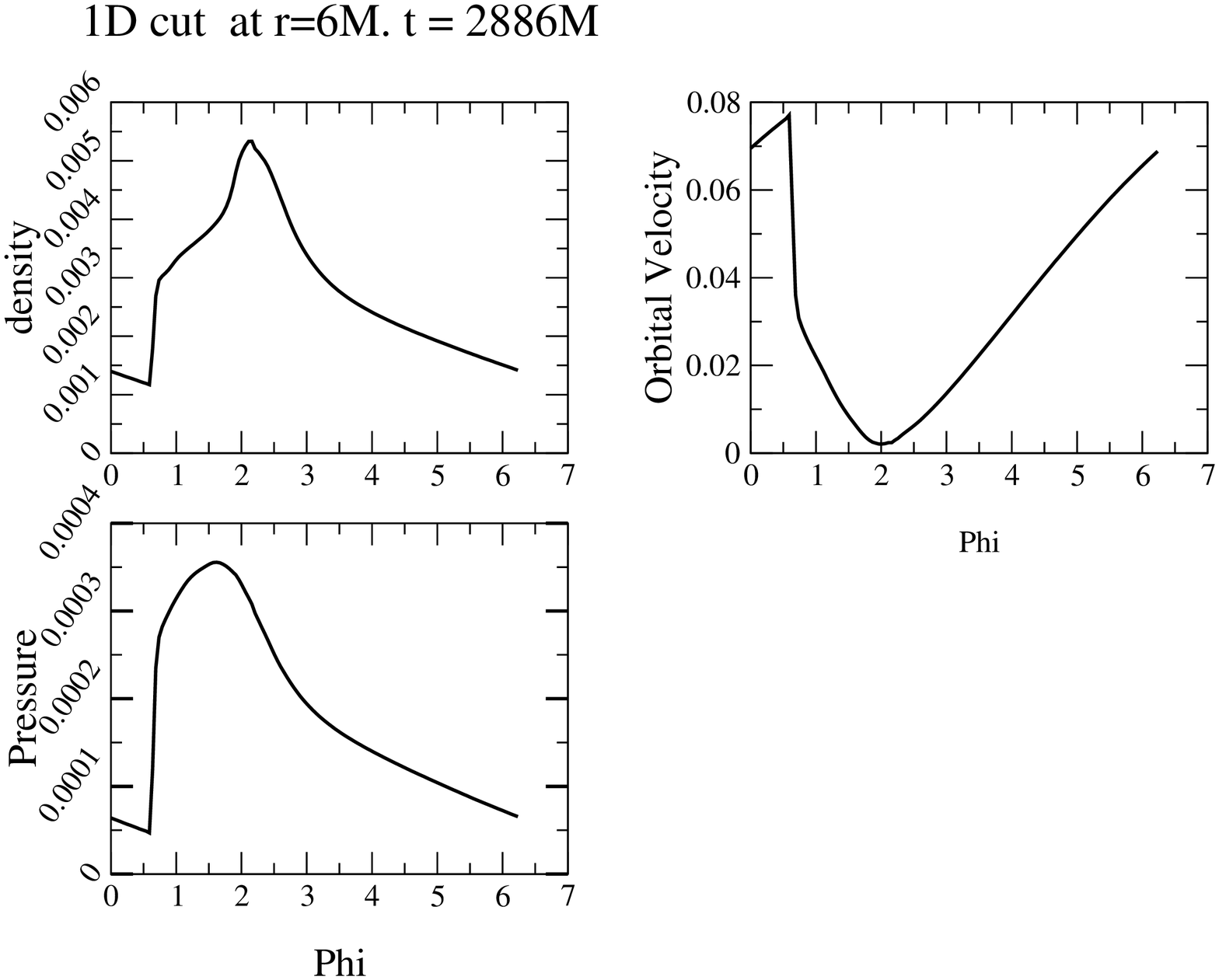}}
\caption{The density, orbital velocity and pressure
vs. angular direction for the
grounded star model at $r=6M$ in 1D. It is  plotted  at $t=2886M$. Strong shock appears 
and it is moving around the black hole.} 
\label{falling star 4}
\end{figure}
\end{center}

The shock wave created by the perturbation destroys the accretion disk. 
The spiral shock wave transports the angular momentum out of disk  when
the matter hits the spiral shock and it causes the
gas falling into the black hole. The falling gas 
does not stop  until reaching  the black hole and the accretion disk
is eventually destroyed by perturbation. Fig. \ref{falling star
3} depicts mass of accretion disk during the numerical evolution. 
As can be seen from this figure, the mass
increases after star is grounded down and then it exponentially decreases during the
evolution. The $1D$ cuts of density, orbital velocity and pressure 
vs. angular coordinate  are shown at fixed radial distance $r=6M$, Fig \ref{falling star 4}. 
The amplitude of shock created on accretion disk in the pressure and 
density plot is greater or equal to a factor of four. 
It is called the strong shock, converting the most of bulk kinetic energy to radiation and 
the thermal energy in the post shock material. Due to the shock waves, angular 
momentum and structure of disk is significantly changed. These parameters 
illustrate that moving shock wave is created on the accretion disk.  The moving shock heats up 
material. We do not expect significant fraction of heated material which could have 
much effect on the overall dynamics of accretion disk. As a results of strong shock  
the matter loses its angular momentum and it begins to fall toward the black hole and 
a large amount of energy is released.

\begin{center}
\begin{figure}
\centerline{
\epsfxsize=12.cm \epsfysize=12.cm
\epsffile{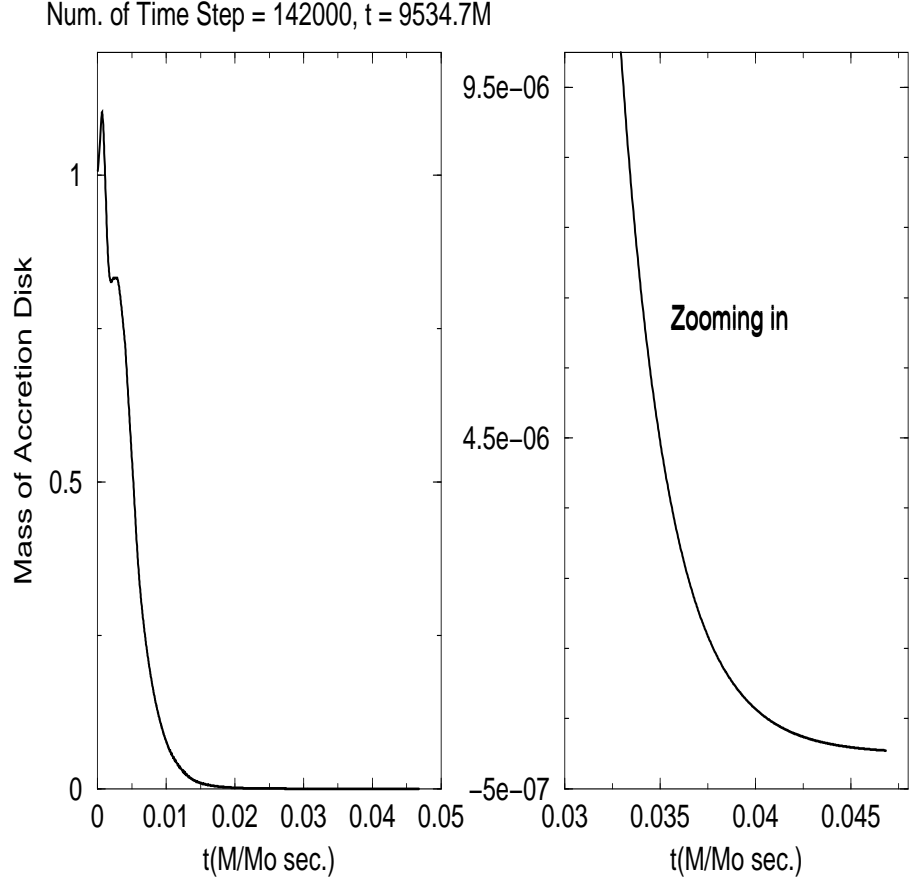}}
\caption{The normalized  mass of accretion disk during the time evolutions for grounded 
star case. Left panel shows the mass of disk during the hole simulation. The peak represents
mass of disk in the moment of star captured. Spiral shock wave can act as a local sink for 
angular momentum and mass of disk decreases drastically, clearly seen in right panel 
(zoom of left panel around the small mass).This comes up with vanishing of accretion disk} 
\label{falling star 3}
\end{figure}
\end{center}

%%%%%%%%%%%%%%%%%%%%%%%%%%%%%%%%%%%%%%%%%%%%%%%%%%%%%%%%%%%%%%%%%%%%%%%%%%
\subsection{Star Makes a Gap on an Accretion Disk}
\label{Star Makes a Hole on an Accretion Disk}

As a second problem, the interaction of a gap with an 
accretion disk is modeled. When the star rotates in their
eccentric orbit, it passes through disk and  creates a gap in the accretion disk. It is 
theoretically explained by Syer et. al. (1991) and Subr et. al. (2004).
In order to model this problem, the same initial steady state accretion disk
is used with previous model, explained in \ref{Modeling of Steady State Initial Disk}. 

\begin{center}
\begin{figure}
\centerline{\epsfxsize=12.cm \epsfysize=12.cm
\epsffile{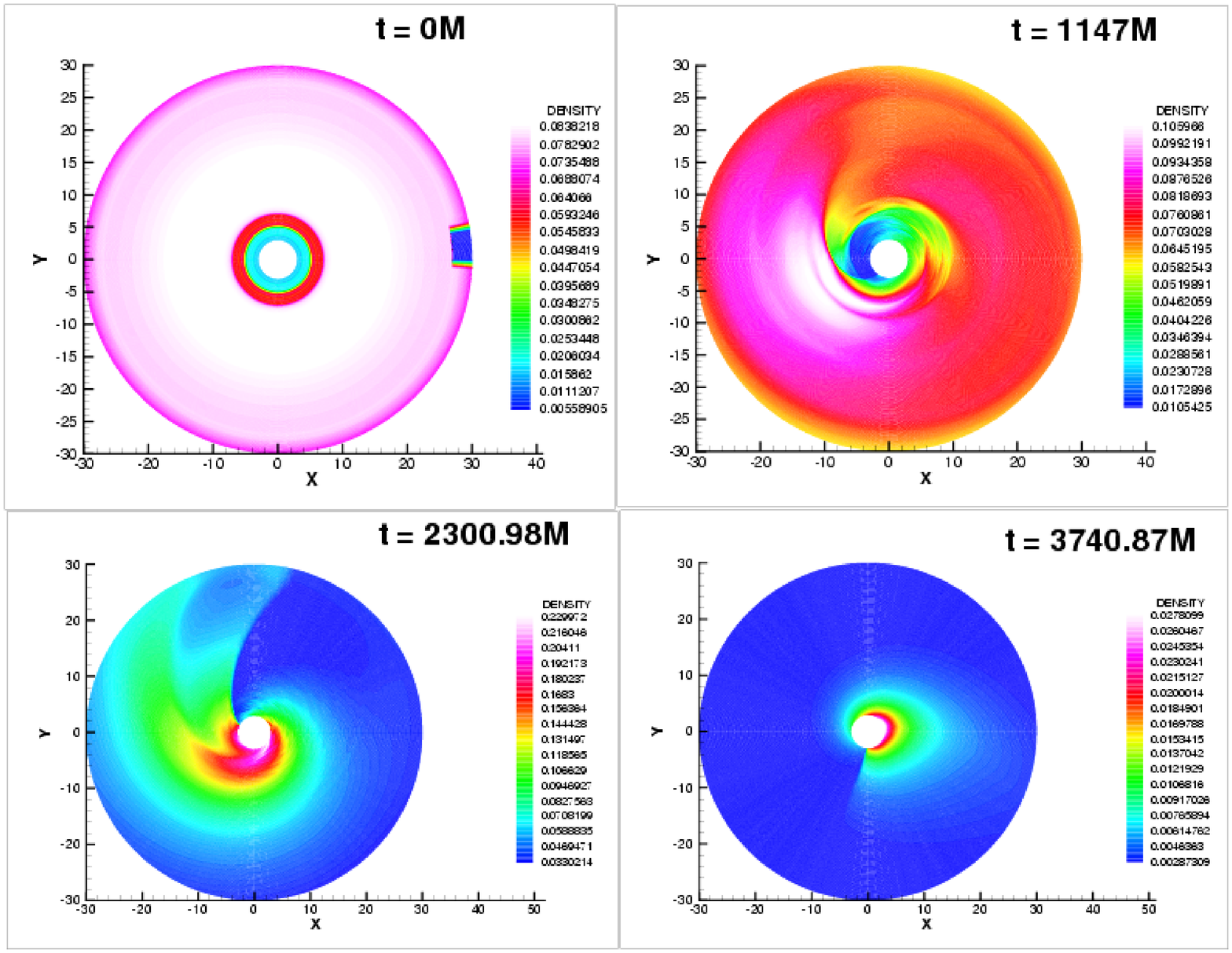}}
\caption{The plots represent four subsequent moments of evolution of perturbed disk by gap
on the $r-\phi$ plane. The color corresponds the counter 
of density $\rho$. The gap is located at close 
to outer boundary, seen at $t=0M$. Initial steady state disk is perturbed by gap and disk 
matter  moves in an unstable manner, and heats up accretion disk. As consequence of this 
phenomenon, one armed spiral shock wave is created at the later times. The spiral wave moves 
at the same direction with disk at the early time of simulation. It rotates counter direction 
to disk after $t=2300M$.} 
\label{making hole 1}
\end{figure}
\end{center}

To model gap-disk interaction, gap is created on the initial steady state accretion disk, 
numerically, taken off the matter at $28M \leq r \leq 30M$ and $0 \leq \phi \leq 0.1$. The
removal of disk gas leads to the formation of a surface density depression that is 
continually replenished by the pressure forces and tidal interaction between gap and 
accretion disk. As  consequence of tidal interaction, the spiral shock wave is created, 
seen in Fig. \ref{making hole 1}. 
Fig. \ref{making hole 1} illustrates the dynamical change of $2D$ representation of disk at 
different snapshot.  Once the gap is created on disk closes to outer boundary, matter on the disk
tries to fill out the gap and as a consequence of the filling, unstable manner is presented on the disk.
This unstable manner forms one-armed spiral wave extending from center to the edge of the disk. 
Because of the spiral wave, surface density of disk exponentially decreases and hence, disk will not 
be able to survive.
 
\begin{center}
\begin{figure}
\centerline{
\epsfxsize=12.cm \epsfysize=12.cm
\epsffile{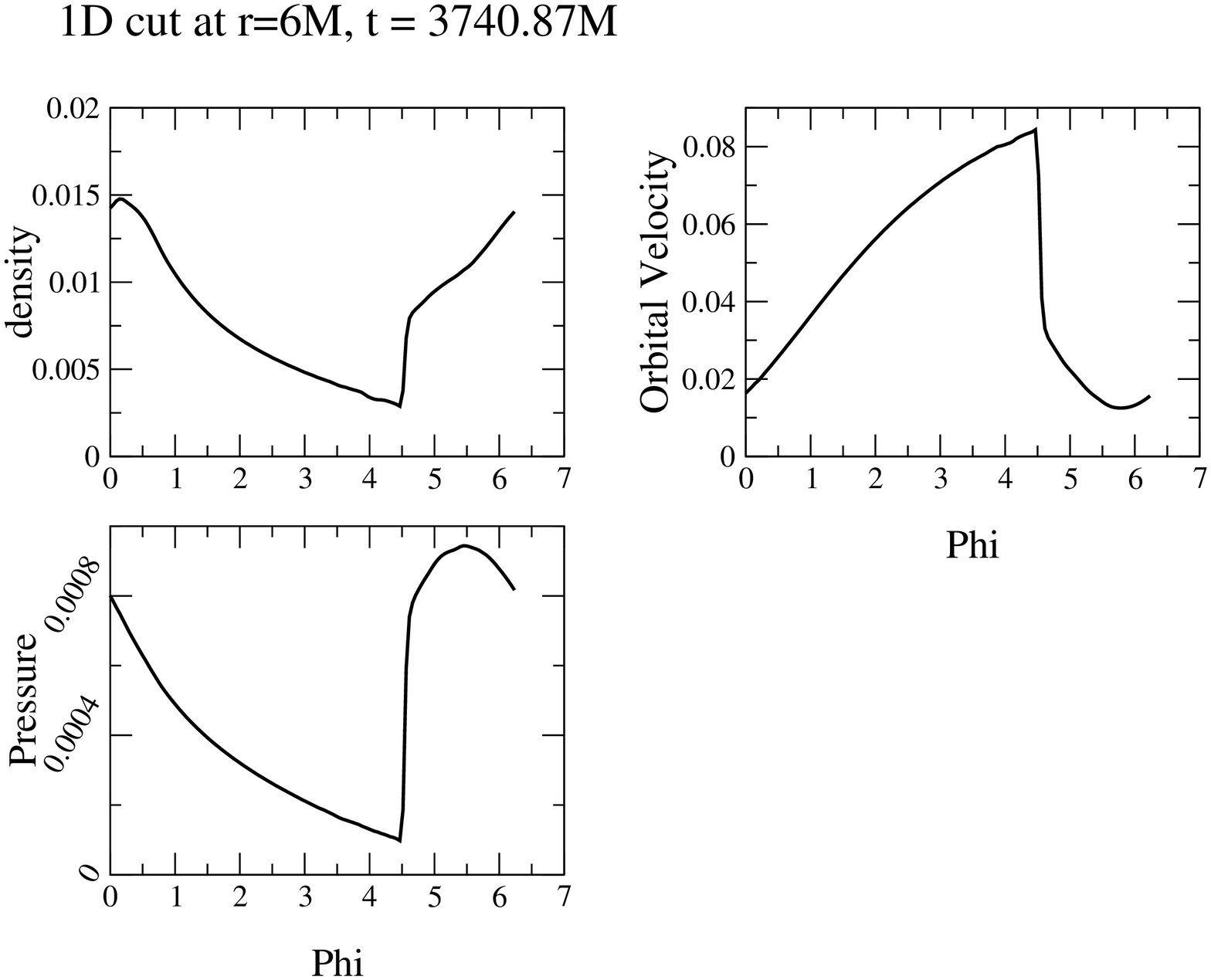}}
\caption{$1D$ representation  of Fig \ref{making hole 1} for 
the density, orbital velocity and pressure
vs. angular direction plotted at $r=6M$ and $t=2886M$. The strong shock 
is presented and dramatic decrease or increase in the parameter causes to gas falling 
into the black  hole.} 
\label{making hole 2}
\end{figure}
\end{center}

Fig. \ref{making hole 2} illustrate the $1D$ cut of density, pressure and orbital velocity
vs. angular coordinate at $r=6M$ and $t=3740.87$. The shock  is numerically 
observed and rotating in the angular direction. 
The produced shock waves   are similar and donate same physical behavior those to which are 
found from the grounded star simulation. These parameters 
illustrate that moving shock wave created on the accretion disk is called the strong 
shock. The shock waves first rotate in the same
direction with accretion disk and then it settles down and counter-rotates
to the disk. The moving shock heats up 
material. We do not expect significant fraction of heated material which could have 
much effect on the overall dynamics of accretion disk. As a results of strong shock,   
the accretion disk loses its angular momentum and the matter begins to fall toward the black 
hole undergoing a transition from subsonic to supersonic speed
when the disk material hits the shock waves.  
The results of perturbing disk by star have same behavior with making the gap on
the accretion disk but the times for creating shock and locations
are occurred at different times and different place, respectively. 
Creation of these phenomena depend on the gravitational, centrifugal, tidal and 
the hydrodynamical forces created on the disk around the black hole. 

%%%%%%%%%%%%%%%%%%%%%%%%%%%%%%%%%%%%%%%%%%%%%%%%%%%%%%%%%%%%%%%%%%%%
\section{CONCLUSION}
We have considered the general relativistic and hydrodynamical effect by
solving general relativistic hydrodynamical equations resulting  from 
perturbing accretion disk around the massive black hole. 
The theoretical predictions have shown that there is an astrophysical 
plausible regime in which an accretion disk around the  massive black hole
in an AGN or other galactic nucleus would dominate the dynamical 
evolution of a star cluster, through the process of star-disk 
interactions. We now discuss the general relativistic astrophysical 
consequences of star-disk interactions.
As should be clear from the numerical simulation  in this paper 
a lot of problems 
still remain in the context of binary formation. Many 
competing explanations for the most basic observations still
exist and it is hard to distinguish between them. But I think with 
the waist effort put into this field, theoretical predictions will
be extended and better during the years, making it possible 
to distinguish really between the different theories and
their outcome. This is however also needed as clearly, theories 
of binary formation are very important. Both since they
probably must be applied to describe the actual formation of a majority 
of all stars but also because a lot of interesting
phenomena like differences between star forming regions, bending of 
jets and formation of clusters can and must be discussed
in connection with the formation and early evolution of binaries. 

The first numerical modeling results of perturbed disk are given in 
this paper. The star and the gap
disk interaction close binary systems are analyzed theoretically and we 
made numerical modeling of a few of these  predications after star opens a gap
on an accretion disk or captured by an accretion disk. The numerical results 
show that regardless of whether stars is ground down or makes a
gap, the star-disk interaction destroys the steady state disk around 
the Schwarzschild black hole. During this process all
the matters at accretion disk fall into the black hole at the center and the 
black hole is getting supermassive. It is the
important physical results. This may explain how the black holes are massive, 
$10^6 - 10^8 M_{\odot}$, at the  center of galaxy and some of the mysteries
in the accretion disk observations. Destroying an accretion disk and making 
massive  compact objects (the black hole and the neutron star) is also declared by
Rauch (1995). Although we star with steady state accretion disk to simulate 
this problem, non-steady state accretion disk case also gives same physical 
and dynamical results.

We stress that the above scenario seems to a remarkable degree model dependent, 
once density of star, size of captured star, initial angular velocity, location of 
perturbation in radial distance, and  how often star can captured by accretion 
disk are specified. Our work may help to understand how the black hole at the AGN could be 
massive and why AGN is bright.

In the star-disk and gap-disk interaction simulations, we have run the
code on Cray T3E machine and Beowulf cluster paralyzing code using SCHMEM and MPI 
libraries.

%%%%%%%%%%%%%%%%%%%%%%%%%%%%%%%%%%%%%%%%%%%%%%%%%%%%%%%%%%%%%%%%%%%%

\begin{acknowledgments}
I would like to thank Joan M. Centrella and Demos Kazanas for a 
useful discussion. This project was carried out 
at NASA/GSFC, Laboratory of High Energy Astrophysics.
It is supported by NASA/GSFC IR\&D. It has been performed using NASA  and TUBITAK/ULAKBIM super 
computers/Beowulf and T3E clusters.
\end{acknowledgments}

\end{document}